\documentclass[prc,preprint,superscriptaddress,floatfix,showpacs,amsmath,amssymb]{revtex4}

\usepackage{graphicx}
\usepackage{dcolumn}
\usepackage{bm}

\newcommand{\phol}{P_{L}^{\gamma}}
\newcommand{\phoc}{P_{c}^{\gamma}}
\newcommand{\tarx}{P_{x}^{T}}
\newcommand{\tary}{P_{y}^{T}}
\newcommand{\tarz}{P_{z}^{T}}
\newcommand{\recx}{P_{x'}^{R}}
\newcommand{\recy}{P_{y'}^{R}}
\newcommand{\recz}{P_{z'}^{R}}
\newcommand{\fpf}[2]{F_{#1}^{*}F_{#2}}
\newcommand{\upa}{\textit{unp}}

\newcommand{\ket}[1]{|{#1}\rangle}
\newcommand{\bra}[1]{\langle{#1}|}

\begin{document}

\title{Calculations of Polarization Observables in Pseudoscalar Meson 
Photo-production Reactions}

\author{A.~M.~Sandorfi}\email[Corresponding author: ]{sandorfi@jlab.org}
\affiliation{Thomas Jefferson National Accelerator Facility, Newport News, Virginia 23606, USA}
\author{S.~Hoblit}
\affiliation{Department of Physics, University of Virginia, Charlottesville, Virginia 22901, USA}
\author{H.~Kamano}
\affiliation{Thomas Jefferson National Accelerator Facility, Newport News, Virginia 23606, USA}
\author{T.-S.~H.~Lee}
\affiliation{Physics Division, Argonne National Laboratory, Illinois 60439, USA}
\affiliation{Thomas Jefferson National Accelerator Facility, Newport News, Virginia 23606, USA}

\date{\today}

\begin{abstract}
In preparation for the extraction of pseudoscalar meson photo-production
amplitudes from a new generation of complete experiments, we assemble the
relations between experimental observables and the Chew-Goldberger-Low-Nambu
amplitudes. We present expressions that allow the direct calculation of
matrix elements with arbitrary spin projections and uses these to clarify
sign differences that exist in the literature. Comparing to the MAID and
SAID analysis codes, we have found that the implied definitions of six
double-polarization observables are the negative of what has been used in
comparing to recent experimental data.
\end{abstract}

\maketitle

\section{Introduction}

As a consequence of dynamic chiral symmetry breaking, the Goldstone bosons $(\pi, \eta, K)$
dress the nucleon and alter its spectrum. Not surprisingly, pseudoscalar meson
production has been a powerful tool in studying the spectrum of excited nucleon
states. However, such states are short lived and broad so that above the energy
of the first resonance, the $P_{33} \Delta$(1232), the excitation spectrum is a complicated
overlap of many resonances. Isolating any one and separating it from backgrounds
has been a long-standing problem in the literature.

The spin degrees of freedom in meson photo-production provide signatures of
interfering partial wave strength that are often dramatic and have been useful
for differentiating between models of meson production amplitudes. 
Models that must account for interfering resonance amplitudes and non-resonant 
contributions are often severely challenged by new polarization data. 
Ideally, one would like to partition the problem by first determining 
the amplitudes from experiment, at least to within a phase, and then relying 
upon a model to separate resonances from non-resonant processes. 
Single-pseudoscalar photo-production is described by 4~complex amplitudes 
(two for the spin states of the photon, two for the nucleon target and two 
for the baryon recoil, which parity considerations reduces to a total of 4). 
To avoid ambiguities, it was shown\cite{chaing} that 
 angular distribution measurements of
at least 8 carefully chosen observables at each energy for both proton and
neutron targets must be performed.
While such experimental information has not yet been available, even after 50 years 
of photo-production experiments, a sequence of \textit{complete} experiments 
are now underway at Jefferson Lab \cite{Frost,HDice}, as well as
complementary experiments at the electron facilities in Bonn and Mainz, with
the goal of obtaining a direct determination of the amplitude to within a
phase, for at least a few production channels, notably $K\Lambda$ and possibly $\pi N$.

Our purpose here is to assemble the relations between experimental observables
and the Chew-Goldberger-Low-Nambu (CGLN) amplitudes \cite{chew}, and to clarify sign
differences that exist in the literature. The four CGLN amplitudes can be expressed in 
Cartesian $(F_{i})$, Spherical or Helicity $(H_{i})$, or Transversity $(b_{i})$ representations. 
While the latter two choices afford some theoretical simplifications when predicting 
asymmetries from model amplitudes \cite{barker}, when working in the reverse direction, 
fitting asymmetries to extract amplitudes, such simplifications are largely moot. 
In practice, one expands the amplitudes in multipoles and fits the multipoles directly. 
This both facilitates the search for resonance behavior and allows the use
of full angular distribution data at a fixed energy to constrain angle-independent
quantities. 
(Extracting the four CGLN amplitudes directly would require separate fits
at each angle, along with some way of constraining an arbitrary phase which could
be angle dependent.) 
Here we restrict our considerations to the CGLN $F_{i}$ representation,
which has the simplest decomposition into multipoles \cite{chew}, 
Eqs.~(\ref{eq:mp-f1})-(\ref{eq:mp-f4}) below.

In single-pseudoscalar meson photo-production there are 16 possible observables, the
cross section ($\sigma$), three single polarization asymmetries determined by the beam,
target or recoil polarizations ($\Sigma$,T,P), and three sets of four asymmetries that
depend on the polarization combinations of beam-target (BT), beam-recoil (BR)
and target-recoil (TR), as in \cite{barker}. 
Expressions for at least some of these observables 
in terms of the CGLN $F_{i}$ appear already 
in earlier papers \cite{donn66,donn72}. The available 
complete expressions can be  classified 
into two groups \cite{adel,fasano} and \cite{drech92,knoch}.
In all cases we have found in the literature, the magnitudes of the expressions relating the
CGLN $F_{i}$ to experimental observables are identical, but the signs of some appear to differ.
In particular, sign differences have occurred in double-polarization observables
for which little data have been available until very recently. 
There is also a set of 37 \textit{Fierz} identities interrelating the 16 polarization 
observables, the most complete list being given in \cite{chaing}. 
We have found many of the signs in the expressions of this list to be incompatible 
with either group of papers, \cite{adel,fasano} or \cite{drech92,knoch}.

\begin{figure}[t]
\includegraphics[clip,width=0.67\textwidth]{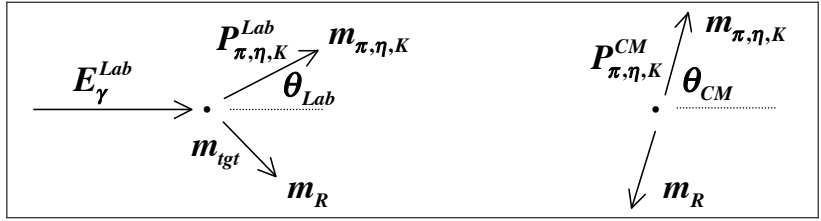}
\caption{\label{fig:kinem} Kinematic variables in meson photo-production in Lab and CM frames.}
\end{figure}

In sections II and III, we give explicit and complete formulae
that allow the direct numerical
calculation of matrix elements with arbitrary spin projections, which are then
used to evaluate polarization observables from
 the four CGLN $F_{i}$ amplitudes. 
In so doing, we resolve the previous sign ambiguities and collect a complete set
of expressions that can be used to extract multipole amplitudes
 from the new set of
\textit{complete} experiments that are now underway. This is 
discussed in section IV with a brief summary.
To avoid possible mis-interpretation of the 
signs of the formulae presented in this paper, the directions of
the polarization vectors for all 16 observables are given explicitly 
in Appendix~\ref{apx_tab}.  The Fierz realtions with 
consistent signs are given in Appendix~\ref{apx_fier}.

\section{Kinematics and Coordinate Definitions}

The kinematic variables of meson photo-production 
used in our derivations are specified in Fig.~\ref{fig:kinem}. 
Some useful relations are :

\begin{itemize}
\item{The total center of mass (CM) energy:
\begin{equation}
W = \sqrt{s}  = \sqrt{ m_{\text{tgt}}(m_{\text{tgt}} + 2E_{\gamma}^{\text{Lab}}) }.
\end{equation} }
\item{The laboratory (Lab) energy needed to excite the hadronic system with total CM energy $W$:
\begin{equation}
E_{\gamma}^{\text{Lab}} = \frac{W^{2} - m_{\text{tgt}}^{2}}{2m_{\text{tgt}}}.
\end{equation} }
\item{The energy of the photon in the CM frame:
\begin{equation}
E_{\gamma}^{\text{CM}}  = \frac{W^{2} - m_{\text{tgt}}^{2}}{2W} = q.
\end{equation} }
\item{The magnitude of the 3-momentum of the meson in the CM frame:
\begin{equation}
\left| P_{\pi ,\eta ,K}^{\text{CM}} \right| = \frac{W}{2}\left\{ {\left[ {1 -
\left( {\frac{{m_{\pi ,\eta ,K}  + m_R }}{W}} \right)^2 }
\right]\left[ {1 - \left( {\frac{{m_{\pi ,\eta ,K}  - m_R }}{W}} \right)^2 } \right]} \right\}^{1/2}
= k.
\end{equation} }
\item{The density of state factor:
\begin{equation}
\rho  = \left| P_{\pi ,\eta ,K}^{\text{CM}} \right| / E_{\gamma}^{cm}.
\end{equation} }
\end{itemize}

\begin{figure}[t]
\includegraphics[clip,width=0.9\textwidth]{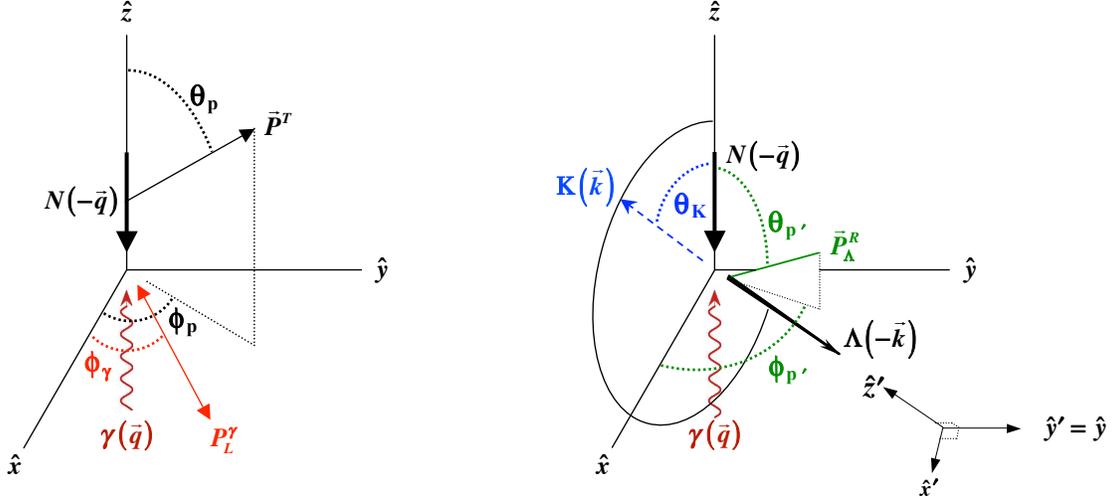}
\caption{\label{fig:coord} The CM coordinate system and angles used to specify polarizations
in the reaction, $\gamma(\vec{q},\vec{P}^{\gamma}) + N(-\vec{q},\vec{P}^{T}) \rightarrow
K(\vec{k}) + \Lambda(-\vec{k},\vec{P}^R_{\Lambda})$. The left (right) side
is for the initial $\gamma N$ (final $K\Lambda$ system. }
\end{figure}

The definitions of Polarization angles used in our derivation
are shown in Fig.~\ref{fig:coord} for the
case of K $\Lambda$ production. The $\langle \hat{x} - \hat{z} \rangle$ plane is
the reaction plane in the center of mass. 
The figure illustrates the case of linear $\gamma$ polarization, with 
the alignment vector $P_{L}^{\gamma}$ (parallel to the oscillating electric 
field of the photon) in the $\langle \hat{x} - \hat{y} \rangle$ plane 
at $\phi_{\gamma}$, rotating clockwise from $\hat{x}$ towards $\hat{y}$. 
The target nucleon polarization $\vec{P}^{T}$ is specified by polar
angle $\theta_{p}$ measured from $\hat{z}$, and azimuthal angle $\phi_{p}$ in
the $\langle \hat{x} - \hat{y} \rangle$ plane, rotating clockwise
from  $\hat{x}$ towards $\hat{y}$. 
The recoil $\Lambda$ is in the $\langle \hat{x} - \hat{z} \rangle$ plane; 
its polarization $\vec{P}_{\Lambda}^{R}$ is at polar $\theta_{p'}$, 
measured from $\hat{z}$, and azimuthal $\phi_{p'}$ in the 
$\langle \hat{x} - \hat{y} \rangle$ plane, rotating clockwise from $\hat{x}$ to $\hat{y}$.

The case of circular photon polarization can potentially lead to some confusion.
Most particle physics literature designates circular states as $r$, for right circular
(or $l$, for left circular), referring to the fact that with $r$ polarization the
electric vector of the photon appears to rotate clockwise \textit{when the photon is
traveling away from the observer}. However, when the same photon is viewed by
an observer facing the incoming photon the electric vector appears to rotate
counter-clockwise. For this reason optics literature traditionally designates
this same state as $l$ circularly polarized. Nonetheless, both conventions agree
on the value of the photon helicity \cite{jackson} $h = \vec{S}\cdot\vec{P} / |\vec{P}| = \pm 1$
and so we use only the helicity designations here, $\vec{P}_{c}^{\gamma} = +1 (-1)$
when 100\% of the photon spins are
parallel (anti-parallel) to the photon momentum vector.

In terms of these polarization vectors, the general form of the cross
section is written as,

\begin{eqnarray}
d\sigma & = & 
\frac{1}{2}\left(
d\sigma_{0} + \hat{\Sigma}[-\phol\cos(2\phi_{\gamma})] + \hat{T}[\tary] + \hat{P}[\recy] 
\right.
\nonumber \\
&& +  \hat{E}[-\phoc\tarz] + \hat{G}[\phol\tarz\sin(2\phi_{\gamma})]
      + \hat{F}[\phoc\tarx] + \hat{H}[\phol\tarx\sin(2\phi_{\gamma})] \nonumber \\
&& +  \hat{C}_{x'}[\phoc\recx] + \hat{C}_{z'}[\phoc\recz]
      + \hat{O}_{x'}[\phol\recx\sin(2\phi_{\gamma})] + \hat{O}_{z'}[\phol\recz\sin(2\phi_{\gamma})] \nonumber \\
&& +  
\left.
\hat{L}_{x'}[\tarz\recx] + \hat{L}_{z'}[\tarz\recz] 
+ \hat{T}_{x'}[\tarx\recx] + \hat{T}_{z'}[\tarx\recz]\right).
\label{eq:general-crs}
\end{eqnarray}

Here $\sigma_{0}$ is the cross section averaged over all 
initial state polarizations, and summed over final state polarization.
We have
designated the product of an asymmetry and $\sigma_{0}$ with a caret, so that $\hat{A} = A\sigma_{0}$.
(These products are referred to as \textit{profile functions} in ref.~\cite{chaing,fasano}.) The terms in
this expression contain the single polarization, BT, BR and TR observables. 
Two terms here have negative coefficients. The first arises because we have taken
for the numerator of the beam asymmetry ($\Sigma$) the somewhat more common definition
of ($\sigma_{\perp} - \sigma_{\parallel}$), rather than the other way around.
(Here $\perp (\parallel)$ corresponds to $\vec P_{\gamma}^{L} = \hat{y} (\hat{x})$ in the
left panel of Fig.~\ref{fig:coord}.)
For the second, because of its connection to sum rules the numerator of the $E$ asymmetry 
is almost always defined in terms of the total CM entrance channel helicity, 
($\sigma_{3/2} - \sigma_{1/2}$), and these correspond to anti-parallel and parallel photon 
and target spin alignments, respectively. 
The specific measurements needed to construct each of these observables is tabulated 
in Appendix~\ref{apx_tab}. (The above expression contains only
the leading polarization terms. Each single polarization observable has another
higher order term that depends on two polarization quantities, and each double
has another term dependent on three polarizations. For simplicity, we defer
these to a later discussion.)

Recoil observables are generally specified in the rotated coordinate system
with $\hat{z}' = +\vec{k}$ (\textit{opposite} to the recoil $\Lambda$ momentum); $\hat{y}' = \hat{y}$
and $\hat{x}' = \hat{y}' \times \hat{z}' = \hat{y} \times \vec{k}$. Occasionally, a particular
recoil observable will have a more transparent interpretation in the unprimed
coordinate system of Fig.~\ref{fig:coord} \cite{CLAS}. Since a baryon polarization transforms
as a standard three vector, the \textit{unprimed} and \textit{primed}
observables are  simply related:
\begin{eqnarray}
A_{x} & = & +A_{x'}\cos\theta_{K}+ A_{z'}\sin\theta_{K} ,\label{eq:br-1} \\
A_{z} & = &  -A_{x'}\sin\theta_{K} + A_{z'}\cos\theta_{K} ,\label{eq:br-2}
\end{eqnarray}
where $A$ represents any one of the BR or TR observables.

\section{\label{sec3} Calculation of polarization observables}

With the variables specified in Fig.~\ref{fig:coord}, 
the differential cross section of 
$\gamma(\vec{q},P^\gamma)+N(-\vec{q},m_{s_N}) \rightarrow 
K(\vec{k})+\Lambda (-\vec{k},m_{s_\Lambda})$
in the center of mass frame can be written as
\begin{eqnarray}
\frac{d\sigma}{d\Omega}(m_{s_\Lambda};P^\gamma,m_{s_N}) =\frac{1}{(4\pi)^2}\frac{k}{q}
\frac{m_Nm_\Lambda}{W^2}
|\bar{u}_\Lambda(-\vec{k},m_{s_\Lambda})I^\mu\epsilon_\mu(q,P^\gamma)
u_N(-\vec{q},m_{s_N})|^2,
\label{eq:dsdo}
\end{eqnarray}
where $W=q+E_N(q)=E_K(k)+E_\Lambda(k)$;
$\epsilon_\mu(q,P^\gamma)$ is the photon polarization vector;
$m_{s_\Lambda}$, and $m_{s_N}$ are the spin quantum number
of the $\Lambda$ and the nucleon in the z-direction, respectively;
$\bar{u}_\Lambda I^\mu\epsilon_\mu u_N$
is normalized to the usual invariant amplitude
calculated from a Lagrangian 
in the convention of Bjorken and Drell~\cite{bjdr}.
The CGLN amplitude is defined by
\begin{eqnarray}
\bar{u}_\Lambda(-\vec{k},m_{s_\Lambda})I^\mu\epsilon_\mu(q,P^\gamma)
 u_N(-\vec{q},m_{s_N}) =-\frac{4\pi W}{\sqrt{m_Nm_\Lambda}}
\chi^+_{m_{s_\Lambda}}F_{CGLN}\chi_{m_{s_N}},
\label{eq:cgln-def}
\end{eqnarray}
where $\chi_{m_s}$ is the usual eigenstate of the Pauli operator $\sigma_z$,
and
\begin{eqnarray}
F_{\text{CGLN}} = \sum_{i=1,4} O_i F_i(\theta_K, E),
\label{eq:cgln}
\end{eqnarray}
with
\begin{eqnarray}
O_1 &=& -i\vec{\sigma}\cdot\vec P^\gamma, \\
O_2 &=&-[\vec{\sigma}\cdot\hat{k}] 
[\vec{\sigma}\cdot (\hat{q}\times \vec P^\gamma)], \\
O_3&=& -i [\vec{\sigma}\cdot\hat{q}][ \hat{k}\cdot\vec P^\gamma], \\
O_4&=&-i [\vec{\sigma}\cdot\hat{k}][\hat{k}\cdot\vec P^\gamma],
\end{eqnarray}
where $\hat{p}=\vec p/|\vec p|$.
We then obtain
\begin{eqnarray}
\frac{d\sigma}{d\Omega}(m_{s_\Lambda};P^\gamma,m_{s_N})=\frac{k}{q}
|\chi^+_{m_{s_\Lambda}}F_{CGLN}\chi_{m_{s_N}}|^2.
\label{eq:dsdo-cgln}
\end{eqnarray}

The formula for calculating CGLN amplitudes from multipoles
are well known~\cite{chew} and are given below:
\begin{eqnarray}
F_1   &=&   \sum_{l=0}^{l_{max}}[
  P_{l+1}'(x) E_{l+} + P_{l-1}'(x)     E_{l-}
+ lP_{l+1}'(x)M_{l+} + (l+1)P_{l-1}'(x) M_{l-}]\,, \label{eq:mp-f1}\\
F_2  & =&    \sum_{l=0}^{l_{max}}[
                    (l+1)P_l'(x)M_{l+} + lP_l'(x) M_{l-}]\,,\label{eq:mp-f2}\\
F_3   &=&    \sum_{l=0}^{l_{max}}[
  P_{l+1}''(x) E_{l+}  + P_{l-1}''(x)   E_{l-}
- P_{l+1}''(x) M_{l+} + P_{l-1}''(x)   M_{l-}]\,,\label{eq:mp-f3}\\
F_4   &=&    \sum_{l=0}^{l_{max}}[
- P_{l}''(x) E_{l+}  - P_{l}''(x)   E_{l-}
+ P_{l}''(x) M_{l+}  - P_{l}''(x)   M_{l-}]\,.\label{eq:mp-f4}
\end{eqnarray}
where $x=\hat{k}\cdot\hat{q} = \cos\theta_K$, $l$ is  the
orbital angular momentum of the $K\Lambda$ system, and $P_{l}'(x)=dP_{l}(x)/dx$ and
$P_{l}''(x)=d^2P_l(x)/dx^2$ are the derivatives of the Legendre function
$P_l(x)$, with the understanding that $P_{-1}'= P_{-1}''=0$.
The value of $l_{max}$ depends obviously on the energy.

In order to calculate the 16 polarization observables
in accordance with
the experimental geometry defined in Appendix~\ref{apx_tab},
we need a formula for calculating cross sections 
with arbitrary spin projections for the initial and final baryon states,
$\gamma(\vec{q}, \vec P^\gamma) + 
N(-\vec{q},\vec P^T) \rightarrow K (\vec{k})
+ \Lambda(-\vec{k},\vec P^R)$, specified in Fig.~\ref{fig:coord}.
Here linear photon polarization must be in
the $\langle\hat x-\hat y\rangle$ plane, but $\vec P^T$ and $\vec P^R$ can be in any directions.
Such cross section formula can be obtained by simply replacing
$|\chi^+_{m_{s_\Lambda}}F_{CGLN}\chi_{m_{s_N}}|^2$ in Eq.~(\ref{eq:dsdo-cgln}) by
$|\bra{\lambda_{\vec P^R}} F_{CGLN}
\ket{\lambda_{\vec P^T}}|^2$:
\begin{eqnarray}
\frac{d\sigma}{d\Omega}(\vec P^\gamma,\vec P^T,\vec P^R)
=\frac{k}{q}
|\bra{\lambda_{\vec P^R}} F_{CGLN}
\ket{\lambda_{\vec P^T}}|^2,
\label{eq:spin-obs}
\end{eqnarray}
where $\ket{\lambda_{\vec P^T}}$ ($\bra{\lambda_{\vec P^R}}$)
is a helicity state of the initial (final) baryon moving in the 
$\vec P^T$ ($\vec P^R$) direction.
The helicity state in the direction
of an arbitrary vector $\vec{p}=(p, \theta_p,\phi_p)$ is defined by
\begin{equation}
\vec S\cdot \hat{p} \ket{\lambda_{\hat{p}}} = \lambda_{\hat{p}}\ket{\lambda_{\hat{p}}},
\label{eq:hel}
\end{equation}
where $\vec S$ is the spin operator. 
For the spin 1/2 particles, $\vec S$ is expressed with
the Pauli matrix: $\vec S = \vec \sigma/2$.

Our next task is to derive  explicit formulae for  calculating
the matrix element $\bra{\lambda_{\vec P^R}} F_{CGLN}
\ket{\lambda_{\vec P^T}}$
in terms of the CGLN amplitudes $F_i$ in Eqs.~(\ref{eq:mp-f1})-(\ref{eq:mp-f4}).
It is known that the helicity state is related to the usual eigenstate
of z-axis quantization by rotations:
\begin{eqnarray}
\ket{\lambda_{\hat{p}}} = \sum_{m_s=\pm 1/2} D^{(1/2)}_{m_s,\lambda_{\hat{p}}}
(\phi_p,\theta_p,-\phi_p) \ket{m_s},
\label{eq:hel-v}
\end{eqnarray}
where $\ket{m_s}$  is defined as $S_z \ket{\pm {1}/{2}}
=(\pm 1/2)\ket{\pm {1}/{2}}$, and
\begin{eqnarray}
D^{(1/2)}_{m_s,\lambda_{\hat p}}(\phi_p,\theta_p,-\phi_p) =e^{-i(m_s-\lambda_{\hat p})\phi_p}
d^{1/2}_{m_s,\lambda_{\hat p}}(\theta_p).
\label{eq:big-d}
\end{eqnarray}
We recall
\begin{eqnarray}
d^{1/2}_{1/2,1/2}(\theta)=d^{1/2}_{-1/2,-1/2}(\theta)=\cos\frac{\theta}{2},
\nonumber \\
d^{1/2}_{-1/2,1/2}(\theta)=-d^{1/2}_{1/2,-1/2}(\theta)=\sin\frac{\theta}{2}.
\label{eq:wig-d}
\end{eqnarray}
Equation~(\ref{eq:hel-v}) can be easily verified by explicit calculations
using the definition (\ref{eq:hel})
and properties~(\ref{eq:big-d}) and~(\ref{eq:wig-d})
for the special cases where $\hat{p} =\hat{x}$, $\hat{y}$, $\hat{z}$ and
the usual definition of the Pauli matrices,
$(\sigma_i)_{mm'}$ ($i=x,y,z$ and 
$m$ (row), $m'$ (column) $=+1/2,-1/2$),
\begin{equation}
\sigma_x= \begin{pmatrix} 0&1\\ 1&0 \end{pmatrix},\
\sigma_y= \begin{pmatrix} 0&-i\\ i&0 \end{pmatrix},\
\sigma_z= \begin{pmatrix} 1&0\\ 0&-1\end{pmatrix}.
\label{eq:pauli}
\end{equation}

From Fig.~\ref{fig:coord}, the momenta and linear photon polarization are expressed as
\begin{eqnarray}
\vec{q}&=& q(0,0,1), \label{eq:q} \\
\vec{k}&=& k(\sin\theta_K, 0, \cos\theta_K), \label{eq:k}\\
\vec P^\gamma&=&(\cos\phi_\gamma, \sin\phi_\gamma, 0). \label{eq:esilon}
\end{eqnarray}
Circular photon polarizations of helicity $\lambda_\gamma$ are 
expressed in terms of linear polarizations as
\begin{equation}
(P_c^\gamma)_{\lambda_\gamma=\pm 1} =
\mp\frac{1}{\sqrt{2}}(P^\gamma_x\pm iP^\gamma_y) .
\label{eq:circular}
\end{equation}
For the initial and final baryon polarizations, we use the spherical variables:
\begin{eqnarray}
\vec P^T&=&(1,\theta_p,\phi_p) \label{eq:p},\\
\vec P^R&=&(1,\theta_{p'},\phi_{p'}) \label{eq:pp}.
\end{eqnarray}

By using Eqs.~(\ref{eq:q})-(\ref{eq:esilon}), we can rewrite
$O_i$ in Eq.~(\ref{eq:cgln}) as
\begin{eqnarray}
O_i =\sum_{n=0,3}C_{i,n}(\theta_K,\phi_\gamma) \sigma_n,
\label{eq:cgln-o}
\end{eqnarray}
where $\sigma_0 = 1$, $\sigma_1=\sigma_x$, $\sigma_2=\sigma_y$,
$\sigma_3=\sigma_z$. 
The explicit form of $C_{i,n}$ is shown in Table~\ref{tab:cin}.

\begin{table}
\caption{$C_{i,n}(\theta_K,\phi_\gamma)$ of 
Eqs.~(\ref{eq:cgln-o}) and~(\ref{eq:spin-mx-1})}
\label{tab:cin}
\begin{ruledtabular}
\begin{tabular}{rrrrr}
   &$n=0$&$n=1$&$n=2$&$n=3$\\ \hline
$i=1$&0&$-i\cos\phi_\gamma$&$-i\sin\phi_\gamma$& 0 \\
$i=2$&$ \sin\theta_K \sin\phi_\gamma$& $i \cos\theta_K \cos\phi_\gamma$& 
$i\cos\theta_K \sin\phi_\gamma$& $-i \sin\theta_K \cos\phi_\gamma$ \\
$i=3$&0&0&0&$-i \sin\theta_K \cos\phi_\gamma$ \\
$i=4$&0&$-i\sin^2\theta_K \cos\phi_\gamma$&
0&$-i\sin\theta_K \cos\theta_K \cos\phi_\gamma$
\end{tabular}
\end{ruledtabular}
\end{table}

By using Eq.~(\ref{eq:hel-v}) and Eqs.~(\ref{eq:cgln}) and~(\ref{eq:cgln-o}),
the photo-production matrix element can then be calculated from
\begin{equation}
\bra{\lambda_{\hat P^R }}F_{\text{CGLN}}\ket{\lambda_{\hat P^T }}
=
\sum_{n=0,3} G_n(\theta_K,\phi_\gamma)\bra{\lambda_{\hat P^R}}\sigma_n\ket{\lambda_{\hat P^T}},
\label{eq:spin-mx}
\end{equation}
with 
\begin{eqnarray}
G_n(\theta_K,\phi_\gamma) &=& 
\sum_{i=1,4} F_i(\theta_K,E) C_{i,n}(\theta_K,\phi_\gamma),
\label{eq:spin-mx-1}\\
\bra{\lambda_{\hat P^R}} \sigma_n \ket{\lambda_{\hat P^T}}&=&
\sum_{m_{s'},m_s=\pm 1/2}D^{(1/2)*}_{m_{s'},\lambda_{\hat P^R }}(\phi_{p'},\theta_{p'},-\phi_{p'})
D^{(1/2)}_{m_s,\lambda_{\hat P^T }}(\phi_p,\theta_{p},-\phi_{p})
\bra{m_{s'}} \sigma_n \ket{m_s},
\label{eq:spin-mx-2}
\end{eqnarray}
where $\bra{m_s'} \sigma_n \ket{m_s}=(\sigma_n)_{m_s',m_s}$ can be calculated from
Eq.~(\ref{eq:pauli}).

We may now generate multipoles from a model and then use 
Eqs.~(\ref{eq:mp-f1})-(\ref{eq:mp-f4}) to calculate the CGLN amplitudes, 
which are then used to calculate the matrix element 
$\bra{\lambda_{\vec P^R}} F_{CGLN} \ket{\lambda_{\vec P^T}}$ by using 
Eqs.~(\ref{eq:spin-mx})-(\ref{eq:spin-mx-2}).
Equation~(\ref{eq:spin-obs}) then allows us 
to calculate all possible  polarization observables.

\section{\label{sec:cgln}Relating Observables to CGLN Amplitudes}

We are now in a position to use any set of multipole amplitudes
to calculate CGLN amplitudes by using Eqs.~(\ref{eq:mp-f1})-(\ref{eq:mp-f4}), and then use these
to evaluate: (a) the polarization observables by using the formulae
described in the previous section and the
spin projections specified in  the table in the Appendix~\ref{apx_tab}, and
(b) the same observables calculated from the analytic expressions,
as given in \cite{adel,fasano} or \cite{drech92,knoch}.
As expected, the absolute magnitudes from two methods
are the same, but some of their signs are different.
In doing so, we are able to fix the signs of the
analytic expressions for the experimental conditions
specified in Fig.~\ref{fig:coord} and Appendix~\ref{apx_tab}. Our results are:
\begin{subequations}
\begin{eqnarray}
\sigma_{0}   & = & + \Re e\left\{ \fpf{1}{1} + \fpf{2}{2} + \sin^{2}\theta(\fpf{3}{3}/2
                   + \fpf{4}{4}/2 + \fpf{2}{3} + \fpf{1}{4} \right. \nonumber \\
             &   &   \mbox{} \left. + \cos\theta\fpf{3}{4}) - 2\cos\theta\fpf{1}{2} \right\} \rho \\
\hat{\Sigma} & = & -\sin^{2}\theta\Re e\left\{ \left(\fpf{3}{3} +\fpf{4}{4}\right)/2 
                   + \fpf{2}{3} + \fpf{1}{4} + \cos\theta\fpf{3}{4}\right\}\rho \\
\hat{T}      & = & + \sin\theta\Im m\left\{\fpf{1}{3} - \fpf{2}{4} + \cos\theta(\fpf{1}{4} - \fpf{2}{3})
                   - \sin^{2}\theta\fpf{3}{4}\right\}\rho \\
\hat{P}      & = & -\sin\theta\Im m\left\{ 2\fpf{1}{2} + \fpf{1}{3} - \fpf{2}{4} - \cos\theta(\fpf{2}{3} -\fpf{1}{4})
                   - \sin^{2}\theta\fpf{3}{4}\right\}\rho \\
\hat{E}      & = & + \Re e\left\{ \fpf{1}{1} + \fpf{2}{2} - 2\cos\theta\fpf{1}{2}
                   + \sin^{2}\theta(\fpf{2}{3} + \fpf{1}{4}) \right\}\rho \\
\hat{G}      & = & + \sin^{2}\theta\Im m\left\{\fpf{2}{3} + \fpf{1}{4}\right\}\rho \\
\hat{F}      & = & + \sin\theta\Re e\left\{\fpf{1}{3} - \fpf{2}{4} - \cos\theta(\fpf{2}{3} - \fpf{1}{4})\right\}\rho \\
\hat{H}      & = & - \sin\theta\Im m\left\{2\fpf{1}{2} + \fpf{1}{3} - \fpf{2}{4}
                   + \cos\theta(\fpf{1}{4} - \fpf{2}{3})\right\}\rho \\
\hat{C}_{x'} & = & - \sin\theta\Re e\left\{\fpf{1}{1} - \fpf{2}{2} - \fpf{2}{3} + \fpf{1}{4} 
                   - \cos\theta(\fpf{2}{4} - \fpf{1}{3})\right\}\rho \\
\hat{C}_{z'} & = & -\Re e\left\{2\fpf{1}{2} - \cos\theta(\fpf{1}{1} + \fpf{2}{2})
                   + \sin^{2}\theta(\fpf{1}{3} + \fpf{2}{4})\right\}\rho \\
\hat{O}_{x'} & = & -\sin\theta\Im m\left\{\fpf{2}{3} - \fpf{1}{4} + \cos\theta(\fpf{2}{4} - \fpf{1}{3})\right\}\rho \\
\hat{O}_{z'} & = & + \sin^{2}\theta\Im m\left\{\fpf{1}{3} + \fpf{2}{4}\right\}\rho 
\end{eqnarray}
\begin{eqnarray}
\hat{L}_{x'} & = & + \sin\theta\Re e\left\{\fpf{1}{1} - \fpf{2}{2} - \fpf{2}{3} + \fpf{1}{4}
                   + \sin^{2}\theta(\fpf{4}{4} - \fpf{3}{3})/2 \right. \nonumber \\
             &   & \mbox{} \left. + \cos\theta(\fpf{1}{3} - \fpf{2}{4})\right\}\rho \\
\hat{L}_{z'} & = & + \Re e\left\{2\fpf{1}{2} - \cos\theta(\fpf{1}{1} + \fpf{2}{2})
                   + \sin^{2}\theta(\fpf{1}{3} + \fpf{2}{4} + \fpf{3}{4}) \right. \nonumber \\
             &   & \mbox{} \left. + \cos\theta\sin^{2}\theta(\fpf{3}{3} + \fpf{4}{4})/2 \right\}\rho \\
\hat{T}_{x'} & = & -\sin^{2}\theta\Re e\left\{\fpf{1}{3} + \fpf{2}{4} + \fpf{3}{4}
                   + \cos\theta(\fpf{3}{3} + \fpf{4}{4})/2 \right\}\rho \\
\hat{T}_{z'} & = & + \sin\theta \Re e \left\{\fpf{1}{4} - \fpf{2}{3}
                   + \cos\theta(\fpf{1}{3} - \fpf{2}{4}) \right. \nonumber \\
             &   & \mbox{} \left. + \sin^{2}\theta(\fpf{4}{4} - \fpf{3}{3})/2 \right\}\rho
\end{eqnarray}
\label{eq:obs-cgln}
\end{subequations}

A comparable set of expressions are given by Fasano, Tabakin and Saghai 
(FTS) in~\cite{fasano}. 
That paper defines the photon polarization using Stokes vectors taken from
optics where right and left circular polarization are interpreted differently.
Nonetheless, they associate photon helicity +1 with what 
Ref.~\cite{fasano} refers to as $r$ circular polarization. 
Keeping this convention and allowing for their different
definition of the $E$ beam-target asymmetry, the above expressions 
are consistent with those of \cite{fasano}.

Comparing the above relations to those given by Kn\"{o}chlein, Drechsel 
and Tiator (KDT) (Appendix B and C of \cite{knoch}), 
six have different signs, the BT observable $H$, the TR observable
$L_{x'}$ and all four of the BR observables 
$C_{x'}$, $C_{z'}$, $O_{x'}$ and $O_{z'}$. 
The KDT paper of \cite{knoch} is listed in the MAID on-line meson production 
analysis \cite{MAID,drech07,mart} as the defining reference 
for the connection between CGLN amplitudes and polarization observables. 
To check if these differences persist in the MAID code we have downloaded 
MAID multipoles, used the relations in Eqs.~(\ref{eq:mp-f1})-(\ref{eq:mp-f4})
to construct from these the four CGLN $F_{i}$ amplitudes, 
and then used our equations~(\ref{eq:obs-cgln})
above to construct observables. Comparing the results to direct predictions of
observables from the MAID code, we find the same six sign differences.
However, in the general form of the cross section given by KDT in~\cite{knoch}
these six observables appear with a negative coefficient, as opposed to
our form of the cross section in Eq.~(\ref{eq:general-crs}).
This is equivalent to interchanging the $\sigma_1$ and $\sigma_2$ measurements
of Appendix~\ref{apx_tab} that are needed to construct these six observables.
The choice of these two measurements that we list in Appendix~\ref{apx_tab}
seem the obvious ones.
They are, with the exception of the $E$ asymmetry, the same choices used by
FTS in~\cite{fasano}.
Despite the fact that KDT refer to their definition of observables as
{\it common} to FTS in~\cite{fasano}, there is evidently a sign difference
for $H$, $L_{x'}$, $C_{x'}$, $C_{z'}$, $O_{x'}$ and $O_{z'}$.

We have conducted a similar test with the GWU/VPI SAID on-line analysis 
code~\cite{SAID,arndt}, downloading SAID multipoles, using the relations 
in Eqs.~(\ref{eq:mp-f1})-(\ref{eq:mp-f4}) to construct from these 
the four CGLN $F_{i}$ amplitudes, 
and then using our equations~(\ref{eq:obs-cgln}) above to construct
observables. When the results are compared to direct predictions of observables
from the SAID code, again the same 6 observables 
($H$, $L_{x'}$ , $C_{x'}$ , $C_{z'}$ , $O_{x'}$ , $O_{z'}$) differ in sign.
For the definition of observables, SAID refers to the Barker, Donnachie
and Storrow paper of~\cite{barker}.
That paper is in general too condensed to definitively address signs, but at
least in the case of $O_{x'}$ they define the required (B,T,R) measurements
as $(\pm\pi/4,-,x')$ which is in agreement with our choice 
in Appendix~\ref{apx_tab}.

New data are emerging from the current generation of polarization experiments
which make these sign differences an important issue.
In Ref.~\cite{CLAS}, recent results for the $C_{x'}$ and $C_{z'}$ asymmetries 
have been compared with the direct predictions of the Kaon-MAID code,
ignoring the sign reversal. 
This has particularly dramatic consequences for the BR asymmetry
$C_{z'}$ which is constrained by angular momentum
conservation to the value of +1 at 0 degrees. 
This is straightforward to see from Appendix~\ref{apx_tab}, where 
$C_{z'} = \{\sigma_{1}(+1,0,+z') - \sigma_{2}(+1,0,-z')\}/\{\sigma_{1} + \sigma_{2}\}$.
When the incident photon spin is oriented along $+\hat{z}$, only
those target nucleons with anti-parallel spin can contribute to the production
of spin zero mesons at 0 degrees, and the projection of the total angular
momentum along  $\hat{z}$ is $+\frac{1}{2}$. 
Thus, the recoil baryon must have its spin oriented
along $+\hat{z} = +\hat{z}'$, so that $\sigma_{2}$ must vanish. 
The recent measurements on $K^{+} \Lambda$ production \cite{CLAS} clearly 
show this asymmetry approaching +1 at
0 degrees, along with MAID predictions approaching $-1$.

\begin{figure}[t]
\includegraphics[clip,width=0.5\textwidth]{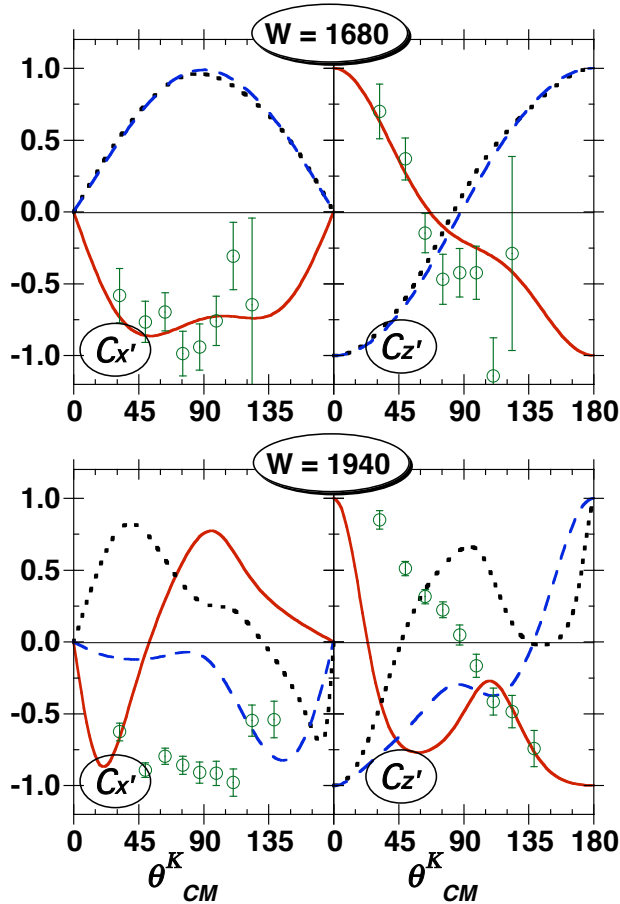}
\caption{\label{fig:cfits} $C_{x'}$ (left) and $C_{z'}$ (right) for the
$\gamma p \rightarrow K^{+}\Lambda$ reaction at $W=1680$ MeV
(top) and $W=1940$ MeV (bottom). Kaon-MAID predictions are dashed \cite{MAID,drech07,mart}, K-SAID
predictions are dotted \cite{SAID,arndt}. Predictions using the multipoles of \cite{bruno} and our
equations are shown as solid red curves. The green circles are from \cite{CLAS}.}
\end{figure}

The trends in $C_{x'}$ and $C_{z'}$ for $\gamma p \rightarrow K^{+}\Lambda$
are illustrated with two energies in Fig.~\ref{fig:cfits}.
The data (green circles) are recent CLAS results from \cite{CLAS} and these are compared
to the direct predictions from Kaon-MAID (blue-dashed curves) and K-SAID (black dotted 
curves) codes. For $C_{z'}$ these clearly have the wrong limits at 0 and 180 degrees. Also
shown are predictions using the multipoles from \cite{bruno} passed through our above
expressions to construct observables (solid red curves). The difference in
signs is also evident in $C_{x'}$ , particularly at low energies where only a
few partial waves are contributing - top panels of Fig.~\ref{fig:cfits}. There it is
clear that the predictions of the different partial solutions are essentially
very similar, differing only in sign.

Finally, we note that when fitting amplitudes from large numbers of measurements
of different observables it will be important to evaluate and treat systematic
uncertainties in a consistent manner. Here the Fierz relations will be
particularly useful, allowing the comparison of data on one observable with
an evaluation in terms of products of other observables. We have numerically
checked the 37 Fierz identities of ref.~\cite{chaing}. 
A revised set with corrected signs is listed in Appendix~\ref{apx_fier}.

In summary, we have explicitly listed the experimental measurements needed to construct
all of the 16 spin observables in pseudoscalar photo-production and provided a consistent
set of equations relating these quantities to the CGLN amplitudes, and from these to
the multipoles.
Comparing to previous works, we have found that the combinations of measurements implied by
the six $H$, $L_{x'}$, $C_{x'}$, $C_{z'}$, $O_{x'}$ and $O_{z'}$ observables calculated
in MAID (and we assume in SAID as well) appears to be the negative of what is in common use
by experimental groups~\cite{CLAS,paterson}.
Neglecting this when fitting amplitudes from complete experiments would drastically alter 
the resultant multipoles.

\begin{acknowledgments}
This work was supported
by the U.S. Department of Energy, Office of Nuclear Physics Division,
under Contract No. DE-AC05-060R23177
under which Jefferson Science Associates operates Jefferson Laboratory,
and also by  the U.S. Department of Energy,
Office of Nuclear Physics Division, under
Contract No. DE-AC02-06CH11357 and
Contract No. DE-FG02-97ER41025.
\end{acknowledgments}

\appendix

\section{\label{apx_tab} Constructing Observables from Measurements}

We tabulate here the pairs of measurements needed to construct each of the
16 transverse photo-production observables in terms of the polarization orientation 
angles of Fig.~\ref{fig:coord}. 
The photon beam is characterized either by its helicity, $h_\gamma$ for circular polarization,
or by $\phi_\gamma^L$ for linear polarization. 
Assuming 100\% polarizations, each observable $\hat A = A\sigma_0 $
is determined by a pair of
measurements, each denoted as $\sigma(B,T,R)$; ``\upa'' indicates the
need to average over the initial spin states of the target or beam, and
to sum over the final spin states of the recoil baryon.
For observables involving only beam or target polarizations,
$\hat{A} = (1/2)(\sigma_{1} - \sigma_{2})$.
For observables involving the final state recoil polarization,
$\hat{A} = (\sigma_{1} - \sigma_{2})$.

\begin{table}[ht]
\caption{\label{tab:1pol} The cross section and observables
involving only one polarization.}
\begin{tabular}{|c|c||c|c||c|c||c|c|} \hline
\multicolumn{2}{|c||}{$\Sigma, T, P$} & \multicolumn{2}{c||}{Beam} &
\multicolumn{2}{c||}{Target} & \multicolumn{2}{c|}{Recoil} \\ \hline
Observable & $(\sigma_{1},\sigma_{2})$ & $h_{\gamma}$ & $\phi_{\gamma}^{L}$ &
$\theta_{p}$ & $\phi_{p}$ & $\theta_{p'}$ & $\phi_{p'}$ \\ \hline
$\sigma_{0}$ & & \upa & \upa & \upa & \upa & \upa & \upa \\ \hline\hline
$\hat{\Sigma}$ & $\sigma_{1} = \sigma(\perp,0,0)$ & - & $\pi/2$ & \upa & \upa & \upa & \upa \\ \cline{2-8}
 & $\sigma_{2} = \sigma(\parallel,0,0)$ & - & 0 & \upa & \upa & \upa & \upa \\ \hline\hline
$\hat{T}$ & $\sigma_{1} = \sigma(0,+y,0)$ & \upa & \upa & $\pi/2$ & $\pi/2$ & \upa & \upa \\ \cline{2-8}
 & $\sigma_{2} = \sigma(0,-y,0)$ & \upa & \upa & $\pi/2$ & $3\pi/2$ & \upa & \upa \\ \hline\hline
$\hat{P}$ & $\sigma_{1} = \sigma(0,0,+y')$ & \upa & \upa & \upa & \upa & $\pi/2$ & $\pi/2$ \\ \cline{2-8}
& $\sigma_{2} = \sigma(0,0,-y')$ & \upa & \upa & \upa & \upa & $\pi/2$ & $3\pi/2$ \\ \hline
\end{tabular}
\end{table}

\begin{table}[ht]
\caption{\label{tab:btpol} Observables involving both beam and target polarization.}
\begin{tabular}{|c|c||c|c||c|c||c|c|} \hline
\multicolumn{2}{|c||}{$B-T$} & \multicolumn{2}{c||}{Beam} &
\multicolumn{2}{c||}{Target} & \multicolumn{2}{c|}{Recoil} \\ \hline
Observable & $(\sigma_{1},\sigma_{2})$ & $h_{\gamma}$ & $\phi_{\gamma}^{L}$ &
$\theta_{p}$ & $\phi_{p}$ & $\theta_{p'}$ & $\phi_{p'}$ \\ \hline
$\hat{E}$ & $\sigma_{1} = \sigma(+1,-z,0)$ & +1 & - & $\pi$ & 0 & \upa & \upa \\ \cline{2-8}
 & $\sigma_{2} = \sigma(+1,+z,0)$ & +1 & - & 0 & 0 & \upa & \upa \\ \hline
$\hat{E}$ & $\sigma_{1} = \sigma(+1,-z,0)$ & +1 & - & $\pi$ & 0 & \upa & \upa \\ \cline{2-8}
 & $\sigma_{2} = \sigma(-1,-z,0)$ & -1 & - & $\pi$ & 0 & \upa & \upa \\ \hline\hline
$\hat{G}$ & $\sigma_{1} = \sigma(+\pi/4,+z,0)$ & - & $\pi/4$ & 0 & 0 & \upa & \upa \\ \cline{2-8}
 & $\sigma_{2} = \sigma(+\pi/4,-z,0)$ & - & $\pi/4$ & $\pi$ & 0 & \upa & \upa \\ \hline
$\hat{G}$ & $\sigma_{1} = \sigma(+\pi/4,+z,0)$ & - & $\pi/4$ & 0 & 0 & \upa & \upa \\ \cline{2-8}
 & $\sigma_{2} = \sigma(-\pi/4,+z,0)$ & - & $3\pi/4$ & 0 & 0 & \upa & \upa \\ \hline\hline
$\hat{F}$ & $\sigma_{1} = \sigma(+1,+x,0)$ & +1 & - & $\pi/2$ & 0 & \upa & \upa \\ \cline{2-8}
 & $\sigma_{2} = \sigma(-1,+x,0)$ & -1 & - & $\pi/2$ & 0 & \upa & \upa \\ \hline
$\hat{F}$ & $\sigma_{1} = \sigma(+1,+x,0)$ & +1 & - & $\pi/2$ & 0 & \upa & \upa \\ \cline{2-8}
 & $\sigma_{2} = \sigma(+1,-x,0)$ & +1 & - & $\pi/2$ & $\pi$ & \upa & \upa \\ \hline\hline
$\hat{H}$ & $\sigma_{1} = \sigma(+\pi/4,+x,0)$ & - & $\pi/4$ & $\pi/2$ & 0 & \upa & \upa \\ \cline{2-8}
 & $\sigma_{2} = \sigma(-\pi/4,+x,0)$ & - & $3\pi/4$ & $\pi/2$ & 0 & \upa & \upa \\ \hline
$\hat{H}$ & $\sigma_{1} = \sigma(+\pi/4,+x,0)$ & - & $\pi/4$ & $\pi/2$ & 0 & \upa & \upa \\ \cline{2-8}
 & $\sigma_{2} = \sigma(+\pi/4,-x,0)$ & - & $\pi/4$ & $\pi/2$ & $\pi$ & \upa & \upa \\ \hline
\end{tabular}
\end{table}

\begin{table}[ht]
\caption{\label{tab:brpol} Observables involving both beam and recoil polarization.}
\begin{tabular}{|c|c||c|c||c|c||c|c|} \hline
\multicolumn{2}{|c||}{$B-R$} & \multicolumn{2}{c||}{Beam} &
\multicolumn{2}{c||}{Target} & \multicolumn{2}{c|}{Recoil} \\ \hline
Observable & $(\sigma_{1},\sigma_{2})$ & $h_{\gamma}$ & $\phi_{\gamma}^{L}$ &
$\theta_{p}$ & $\phi_{p}$ & $\theta_{p'}$ & $\phi_{p'}$ \\ \hline
$\hat{C}_{x'}$ & $\sigma_{1} = \sigma(+1,0,+x')$ & +1 & - & \upa & \upa & $\pi/2 + \theta_{K}$ & 0 \\ \cline{2-8}
 & $\sigma_{2} = \sigma(-1,0,+x')$ & -1 & - & \upa & \upa & $\pi/2 + \theta_{K}$ & 0 \\ \hline
$\hat{C}_{x'}$ & $\sigma_{1} = \sigma(+1,0,+x')$ & +1 & - & \upa & \upa & $\pi/2 + \theta_{K}$ & 0 \\ \cline{2-8}
 & $\sigma_{2} = \sigma(+1,0,-x')$ & +1 & - & \upa & \upa & $3\pi/2 + \theta_{K}$ & 0 \\ \hline\hline
$\hat{C}_{z'}$ & $\sigma_{1} = \sigma(+1,0,+z')$ & +1 & - & \upa & \upa & $\theta_{K}$ & 0 \\ \cline{2-8}
 & $\sigma_{2} = \sigma(-1,0,+z')$ & -1 & - & \upa & \upa & $\theta_{K}$ & 0 \\ \hline
$\hat{C}_{z'}$ & $\sigma_{1} = \sigma(+1,0,+z')$ & +1 & - & \upa & \upa & $\theta_{K}$ & 0 \\ \cline{2-8}
 & $\sigma_{2} = \sigma(+1,0,-z')$ & +1 & - & \upa & \upa & $\pi + \theta_{K}$ & 0 \\ \hline\hline
$\hat{O}_{x'}$ & $\sigma_{1} = \sigma(+\pi/4,0,+x')$ & - & $\pi/4$ & \upa & \upa & $\pi/2 + \theta_{K}$ & 0 \\ \cline{2-8}
 & $\sigma_{2} = \sigma(-\pi/4,0,+x')$ & - & $3\pi/4$ & \upa & \upa & $\pi/2 + \theta_{K}$ & 0 \\ \hline
$\hat{O}_{x'}$ & $\sigma_{1} = \sigma(+\pi/4,0,+x')$ & - & $\pi/4$ & \upa & \upa & $\pi/2 + \theta_{K}$ & 0 \\ \cline{2-8}
 & $\sigma_{2} = \sigma(+\pi/4,0,-x')$ & - & $\pi/4$ & \upa & \upa & $3\pi/2 + \theta_{K}$ & 0 \\ \hline\hline
$\hat{O}_{z'}$ & $\sigma_{1} = \sigma(+\pi/4,0,+z')$ & - & $\pi/4$ & \upa & \upa & $\theta_{K}$ & 0 \\ \cline{2-8}
 & $\sigma_{2} = \sigma(-\pi/4,0,+z')$ & - & $3\pi/4$ & \upa & \upa & $\theta_{K}$ & 0 \\ \hline
$\hat{O}_{z'}$ & $\sigma_{1} = \sigma(+\pi/4,0,+z')$ & - & $\pi/4$ & \upa & \upa & $\theta_{K}$ & 0 \\ \cline{2-8}
 & $\sigma_{2} = \sigma(+\pi/4,0,-z')$ & - & $\pi/4$ & \upa & \upa & $\pi + \theta_{K}$ & 0 \\ \hline
\end{tabular}
\end{table}

\begin{table}[ht]
\caption{\label{tab:trpol} Observables involving both target and recoil polarization.}
\begin{tabular}{|c|c||c|c||c|c||c|c|} \hline
\multicolumn{2}{|c||}{$T-R$} & \multicolumn{2}{c||}{Beam} &
\multicolumn{2}{c||}{Target} & \multicolumn{2}{c|}{Recoil} \\ \hline
Observable & $(\sigma_{1},\sigma_{2})$ & $h_{\gamma}$ & $\phi_{\gamma}^{L}$ &
$\theta_{p}$ & $\phi_{p}$ & $\theta_{p'}$ & $\phi_{p'}$ \\ \hline
$\hat{L}_{x'}$ & $\sigma_{1} = \sigma(0,+z,+x')$ & \upa & \upa & 0 & 0 & $\pi/2 + \theta_{K}$ & 0 \\ \cline{2-8}
 & $\sigma_{2} = \sigma(0,-z,+x')$ & \upa & \upa & $\pi$ & 0 & $\pi/2 + \theta_{K}$ & 0 \\ \hline
$\hat{L}_{x'}$ & $\sigma_{1} = \sigma(0,+z,+x')$ & \upa & \upa & 0 & 0 & $\pi/2 + \theta_{K}$ & 0 \\ \cline{2-8}
 & $\sigma_{2} = \sigma(0,+z,-x')$ & \upa & \upa & 0 & 0 & $3\pi/2 + \theta_{K}$ & 0 \\ \hline\hline
$\hat{L}_{z'}$ & $\sigma_{1} = \sigma(0,+z,+z')$ & \upa & \upa & 0 & 0 & $\theta_{K}$ & 0 \\ \cline{2-8}
 & $\sigma_{2} = \sigma(0,-z,+z')$ & \upa & \upa & $\pi$ & 0 & $\theta_{K}$ & 0 \\ \hline
$\hat{L}_{z'}$ & $\sigma_{1} = \sigma(0,+z,+z')$ & \upa & \upa & 0 & 0 & $\theta_{K}$ & 0 \\ \cline{2-8}
 & $\sigma_{2} = \sigma(0,+z,-z')$ & \upa & \upa & 0 & 0 & $\pi + \theta_{K}$ & 0 \\ \hline\hline
$\hat{T}_{x'}$ & $\sigma_{1} = \sigma(0,+x,+x')$ & \upa & \upa & $\pi/2$ & 0 & $\pi/2 + \theta_{K}$ & 0 \\ \cline{2-8}
 & $\sigma_{2} = \sigma(0,-x,+x')$ & \upa & \upa & $\pi/2$ & $\pi$ & $\pi/2 + \theta_{K}$ & 0 \\ \hline
$\hat{T}_{x'}$ & $\sigma_{1} = \sigma(0,+x,+x')$ & \upa & \upa & $\pi/2$ & 0 & $\pi/2 + \theta_{K}$ & 0 \\ \cline{2-8}
 & $\sigma_{2} = \sigma(0,+x,-x')$ & \upa & \upa & $\pi/2$ & 0 & $3\pi/2 + \theta_{K}$ & 0 \\ \hline\hline
$\hat{T}_{z'}$ & $\sigma_{1} = \sigma(0,+x,+z')$ & \upa & \upa & $\pi/2$ & 0 & $\theta_{K}$ & 0 \\ \cline{2-8}
 & $\sigma_{2} = \sigma(0,-x,+z')$ & \upa & \upa & $\pi/2$ & $\pi$ & $\theta_{K}$ & 0 \\ \hline
$\hat{T}_{z'}$ & $\sigma_{1} = \sigma(0,+x,+z')$ & \upa & \upa & $\pi/2$ & 0 & $\theta_{K}$ & 0 \\ \cline{2-8}
 & $\sigma_{2} = \sigma(0,+x,-z')$ & \upa & \upa & $\pi/2$ & 0 & $\pi + \theta_{K}$ & 0 \\ \hline
\end{tabular}
\end{table}

\clearpage

\section{\label{apx_fier} The Fierz Identities}

We list here the Fierz identities relating \textit{asymmetries}, with corrected
signs. The equation numbering sequence is that of Chiang and Tabakin \cite{chaing}.
Compared to the latter, signs have changed in all but (L.1), (L.4-6),
(Q.r), (Q.bt.3), (Q.tr.1-2), and of course the six \textit{Squared} relations.
Sign changes in eight of the equations can be attributed to the
different definition for the $E$ asymmetry used by Fasano, Tabakin
and Saghai \cite{fasano}, to which Chiang and Tabakin refer. (We note that
sign changes in another 15 equations could have been explained if
the definition of the beam asymmetry ($\Sigma$) were also reversed; but
our definition of $\Sigma$ in Table A1 is identical to that of Fasano,
Tabakin and Saghai \cite{fasano}.)

\subsection{Linear-Quadratic relations}

\begin{equation}
\begin{split}
1 =& \{ \Sigma^{2} + T^{2} + P^{2} + E^{2} + G^{2} + F^{2} + H^{2} \\
  &+ O_{x'}^{2} + O_{z'}^{2} + C_{x'}^{2} + C_{z'}^{2} 
  + L_{x'}^{2} + L_{z'}^{2} + T_{x'}^{2} + T_{z'}^{2}\}/3
\end{split}
\tag{L.0}\label{eq:L.0}
\end{equation}
\begin{equation}
\Sigma = + TP + T_{x'}L_{z'} - T_{z'}L_{x'}
\tag{L.TR}\label{eq:L.TR}
\end{equation}
\begin{equation}
T = + \Sigma P - C_{x'}O_{z'} + C_{z'}O_{x'}
\tag{L.BR}\label{eq:L.BR}
\end{equation}
\begin{equation}
P = + \Sigma T + GF + EH
\tag{L.BT}\label{eq:L.BT}
\end{equation}
\begin{equation}
G = + PF + O_{x'}L_{x'} + O_{z'}L_{z'}
\tag{L.1}\label{eq:L.1}
\end{equation}
\begin{equation}
H = + PE + O_{x'}T_{x'} + O_{z'}T_{z'}
\tag{L.2}\label{eq:L.2}
\end{equation}
\begin{equation}
E = + PH - C_{x'}L_{x'} - C_{z'}L_{z'}
\tag{L.3}\label{eq:L.3}
\end{equation}
\begin{equation}
F = + PG + C_{x'}T_{x'} + C_{z'}T_{z'}
\tag{L.4}\label{eq:L.4}
\end{equation}
\begin{equation}
O_{x'} = + TC_{z'} + GL_{x'} + HT_{x'}
\tag{L.5}\label{eq:L.5}
\end{equation}
\begin{equation}
O_{z'} = - TC_{x'} + GL_{z'} + HT_{z'}
\tag{L.6}\label{eq:L.6}
\end{equation}
\begin{equation}
C_{x'} = - TO_{z'} - EL_{x'} + FT_{x'}
\tag{L.7}\label{eq:L.7}
\end{equation}
\begin{equation}
C_{z'} = + TO_{x'} - EL_{z'} + FT_{z'}
\tag{L.8}\label{eq:L.8}
\end{equation}
\begin{equation}
T_{x'} = + \Sigma L_{z'} + HO_{x'} + FC_{x'}
\tag{L.9}\label{eq:L.9}
\end{equation}
\begin{equation}
T_{z'} = - \Sigma L_{x'} + HO_{z'} + FC_{z'}
\tag{L.10}\label{eq:L.10}
\end{equation}
\begin{equation}
L_{x'} = - \Sigma T_{z'} + GO_{x'} - EC_{x'}
\tag{L.11}\label{eq:L.11}
\end{equation}
\begin{equation}
L_{z'} = + \Sigma T_{x'} + GO_{z'} - EC_{z'}
\tag{L.12}\label{eq:L.12}
\end{equation}

\subsection{Quadratic relations}
\begin{equation}
C_{x'}O_{x'} + C_{z'}O_{z'} + EG - FH = 0
\tag{Q.b}\label{eq:Q.b}
\end{equation}
\begin{equation}
GH - EF - L_{x'}T_{x'} - L_{z'}T_{z'} = 0
\tag{Q.t}\label{eq:Q.t}
\end{equation}
\begin{equation}
C_{x'}C_{z'} + O_{x'}O_{z'} - L_{x'}L_{z'} - T_{x'}T_{z'} = 0
\tag{Q.r}\label{eq:Q.r}
\end{equation}
\begin{equation}
\Sigma G - TF - O_{z'}T_{x'} + O_{x'}T_{z'} = 0
\tag{Q.bt.1}\label{eq:Q.bt.1}
\end{equation}
\begin{equation}
\Sigma H - TE + O_{z'}L_{x'} - O_{x'}L_{z'} = 0
\tag{Q.bt.2}\label{eq:Q.bt.2}
\end{equation}
\begin{equation}
\Sigma E - TH + C_{z'}T_{x'} - C_{x'}T_{z'} = 0
\tag{Q.bt.3}\label{eq:Q.bt.3}
\end{equation}
\begin{equation}
\Sigma F - TG + C_{z'}L_{x'} - C_{x'}L_{z'} = 0
\tag{Q.bt.4}\label{eq:Q.bt.4}
\end{equation}
\begin{equation}
\Sigma O_{x'} - PC_{z'} + GT_{z'} - HL_{z'} = 0
\tag{Q.br.1}\label{eq:Q.br.1}
\end{equation}
\begin{equation}
\Sigma O_{z'} + PC_{x'} - GT_{x'} + HL_{x'} = 0
\tag{Q.br.2}\label{eq:Q.br.2}
\end{equation}
\begin{equation}
\Sigma C_{x'} + PO_{z'} - ET_{z'} - FL_{z'} = 0
\tag{Q.br.3}\label{eq:Q.br.3}
\end{equation}
\begin{equation}
\Sigma C_{z'} - PO_{x'} + ET_{x'} + FL_{x'} = 0
\tag{Q.br.4}\label{eq:Q.br.4}
\end{equation}
\begin{equation}
TT_{x'} - PL_{z'} - HC_{z'} + FO_{z'} = 0
\tag{Q.tr.1}\label{eq:Q.tr.1}
\end{equation}
\begin{equation}
TT_{z'} + PL_{x'} + HC_{x'} - FO_{x'} = 0
\tag{Q.tr.2}\label{eq:Q.tr.2}
\end{equation}
\begin{equation}
TL_{x'} + PT_{z'} - GC_{z'} - EO_{z'} = 0
\tag{Q.tr.3}\label{eq:Q.tr.3}
\end{equation}
\begin{equation}
TL_{z'} - PT_{x'} + GC_{x'} + EO_{x'} = 0
\tag{Q.tr.4}\label{eq:Q.tr.4}
\end{equation}

\subsection{Squared relations}
\begin{equation}
G^{2} + H^{2} + E^{2} + F^{2} + \Sigma^{2} + T^{2} - P^{2} = 1
\tag{S.bt}\label{eq:S.bt}
\end{equation}
\begin{equation}
O_{x'}^{2} + O_{z'}^{2} + C_{x'}^{2} + C_{z'}^{2} + \Sigma^{2} - T^{2} + P^{2} = 1
\tag{S.br}\label{eq:S.br}
\end{equation}
\begin{equation}
T_{x'}^{2} + T_{z'}^{2} + L_{x'}^{2} + L_{z'}^{2} - \Sigma^{2} + T^{2} + P^{2} = 1
\tag{S.tr}\label{eq:S.tr}
\end{equation}
\begin{equation}
G^{2} + H^{2} - E^{2} - F^{2} - O_{x'}^{2} - O_{z'}^{2} + C_{x'}^{2} + C_{z'}^{2} = 0
\tag{S.b}\label{eq:S.b}
\end{equation}
\begin{equation}
G^{2} - H^{2} + E^{2} - F^{2} + T_{x'}^{2} + T_{z'}^{2} - L_{x'}^{2} - L_{z'}^{2} = 0
\tag{S.t}\label{eq:S.t}
\end{equation}
\begin{equation}
O_{x'}^{2} - O_{z'}^{2} + C_{x'}^{2} - C_{z'}^{2} - T_{x'}^{2} + T_{z'}^{2} - L_{x'}^{2} + L_{z'}^{2} = 0
\tag{S.r}\label{eq:S.r}
\end{equation}


\begin{thebibliography}{21}
\expandafter\ifx\csname natexlab\endcsname\relax\def\natexlab#1{#1}\fi
\expandafter\ifx\csname bibnamefont\endcsname\relax
  \def\bibnamefont#1{#1}\fi
\expandafter\ifx\csname bibfnamefont\endcsname\relax
  \def\bibfnamefont#1{#1}\fi
\expandafter\ifx\csname citenamefont\endcsname\relax
  \def\citenamefont#1{#1}\fi
\expandafter\ifx\csname url\endcsname\relax
  \def\url#1{\texttt{#1}}\fi
\expandafter\ifx\csname urlprefix\endcsname\relax\def\urlprefix{URL }\fi
\providecommand{\bibinfo}[2]{#2}
\providecommand{\eprint}[2][]{\url{#2}}

\bibitem[{\citenamefont{Chaing and Tabakin}(1997)}]{chaing}
\bibinfo{author}{\bibfnamefont{W.-T.} \bibnamefont{Chaing}} \bibnamefont{and}
  \bibinfo{author}{\bibfnamefont{F.}~\bibnamefont{Tabakin}},
  \bibinfo{journal}{\prc} \textbf{\bibinfo{volume}{55}}, \bibinfo{pages}{2054}
  (\bibinfo{year}{1997}).

\bibitem[{Fro()}]{Frost}
\emph{\bibinfo{title}{The g9-frost series of experiments}},
  \urlprefix\url{http://clasweb.jlab.org/shift/g9/}.

\bibitem[{HDi()}]{HDice}
\emph{\bibinfo{title}{The g14-HDice experiment}},
  \urlprefix\url{http://www.jlab.org/exp_prog/proposals/06/PR-06-101.pdf}.

\bibitem[{\citenamefont{Chew et~al.}(1957)}]{chew}
\bibinfo{author}{\bibfnamefont{G.~F.} \bibnamefont{Chew}} \bibnamefont{et~al.},
  \bibinfo{journal}{Phys. Rev.} \textbf{\bibinfo{volume}{105}},
  \bibinfo{pages}{1345} (\bibinfo{year}{1957}).

\bibitem[{\citenamefont{Barker et~al.}(1975)}]{barker}
\bibinfo{author}{\bibfnamefont{I.~S.} \bibnamefont{Barker}}
  \bibnamefont{et~al.}, \bibinfo{journal}{Nucl. Phys. B}
  \textbf{\bibinfo{volume}{95}}, \bibinfo{pages}{347} (\bibinfo{year}{1975}).

\bibitem[{\citenamefont{Donnachie and Shaw}(1966)}]{donn66}
\bibinfo{author}{\bibfnamefont{A.}~\bibnamefont{Donnachie}} \bibnamefont{and}
  \bibinfo{author}{\bibfnamefont{G.}~\bibnamefont{Shaw}},
  \bibinfo{journal}{\ap} \textbf{\bibinfo{volume}{37}}, \bibinfo{pages}{333}
  (\bibinfo{year}{1966}).

\bibitem[{\citenamefont{Donnachie}(1972)}]{donn72}
\bibinfo{author}{\bibfnamefont{A.}~\bibnamefont{Donnachie}},
  \bibinfo{journal}{Pure and Applied Physics} \textbf{\bibinfo{volume}{25-V}},
  \bibinfo{pages}{1} (\bibinfo{year}{1972}), \bibinfo{note}{ed. E.H.S. Bishop,
  Academic Press, NY}.

\bibitem[{\citenamefont{Adelseck and Saghai}(1990)}]{adel}
\bibinfo{author}{\bibfnamefont{R.~A.} \bibnamefont{Adelseck}} \bibnamefont{and}
  \bibinfo{author}{\bibfnamefont{B.}~\bibnamefont{Saghai}},
  \bibinfo{journal}{\prc} \textbf{\bibinfo{volume}{42}}, \bibinfo{pages}{108}
  (\bibinfo{year}{1990}).

\bibitem[{\citenamefont{Fasano et~al.}(1992)\citenamefont{Fasano, Tabakin, and
  Saghai}}]{fasano}
\bibinfo{author}{\bibfnamefont{C.~G.} \bibnamefont{Fasano}},
  \bibinfo{author}{\bibfnamefont{F.}~\bibnamefont{Tabakin}}, \bibnamefont{and}
  \bibinfo{author}{\bibfnamefont{B.}~\bibnamefont{Saghai}},
  \bibinfo{journal}{\prc} \textbf{\bibinfo{volume}{46}}, \bibinfo{pages}{2430}
  (\bibinfo{year}{1992}).

\bibitem[{\citenamefont{Drechsel and Tiator}(1992)}]{drech92}
\bibinfo{author}{\bibfnamefont{D.}~\bibnamefont{Drechsel}} \bibnamefont{and}
  \bibinfo{author}{\bibfnamefont{L.}~\bibnamefont{Tiator}},
  \bibinfo{journal}{J.\ Phys.\ G: Nucl.\ Part.\ Phys.}
  \textbf{\bibinfo{volume}{18}}, \bibinfo{pages}{449} (\bibinfo{year}{1992}).

\bibitem[{\citenamefont{Knochlein et~al.}(1995)\citenamefont{Knochlein,
  Drechsel, and Tiator}}]{knoch}
\bibinfo{author}{\bibfnamefont{G.}~\bibnamefont{Knochlein}},
  \bibinfo{author}{\bibfnamefont{D.}~\bibnamefont{Drechsel}}, \bibnamefont{and}
  \bibinfo{author}{\bibfnamefont{L.}~\bibnamefont{Tiator}},
  \bibinfo{journal}{Z.\ Phys.\ A} \textbf{\bibinfo{volume}{352}},
  \bibinfo{pages}{327} (\bibinfo{year}{1995}).

\bibitem[{\citenamefont{Jackson}(1975)}]{jackson}
\bibinfo{author}{\bibfnamefont{J.~D.} \bibnamefont{Jackson}},
  \emph{\bibinfo{title}{Classical Electrodynamics}} (\bibinfo{publisher}{John
  Wiley \& Sons, NY}, \bibinfo{year}{1975}).

\bibitem[{\citenamefont{Collaboration et~al.}(2007)\citenamefont{Collaboration,
  Bradford et~al.}}]{CLAS}
\bibinfo{author}{\bibfnamefont{CLAS}~\bibnamefont{Collaboration}},
  \bibinfo{author}{\bibfnamefont{R.~K.} \bibnamefont{Bradford}},
  \bibnamefont{et~al.}, \bibinfo{journal}{\prc} \textbf{\bibinfo{volume}{75}},
  \bibinfo{pages}{035205} (\bibinfo{year}{2007}).

\bibitem[{\citenamefont{Bjorken and Drell}(1964)}]{bjdr}
\bibinfo{author}{\bibfnamefont{J.~D.} \bibnamefont{Bjorken}} \bibnamefont{and}
  \bibinfo{author}{\bibfnamefont{S.~D.} \bibnamefont{Drell}},
  \emph{\bibinfo{title}{Relativistic Quantum Mechanics}}
  (\bibinfo{publisher}{McGraw-Hill}, \bibinfo{address}{New York},
  \bibinfo{year}{1964}).

\bibitem[{MAI()}]{MAID}
\emph{\bibinfo{title}{MAID and Kaon-MAID isobar models of meson production}},
  \urlprefix\url{http://wwwkph.kph.uni-mainz.de/MAID/}.

\bibitem[{\citenamefont{Drechsel et~al.}(2007)\citenamefont{Drechsel, Kamalov,
  and Tiator}}]{drech07}
\bibinfo{author}{\bibfnamefont{D.}~\bibnamefont{Drechsel}},
  \bibinfo{author}{\bibfnamefont{S.~S.} \bibnamefont{Kamalov}},
  \bibnamefont{and} \bibinfo{author}{\bibfnamefont{L.}~\bibnamefont{Tiator}},
  \bibinfo{journal}{Eur.\ Phys.\ J.\ A} \textbf{\bibinfo{volume}{34}},
  \bibinfo{pages}{69} (\bibinfo{year}{2007}).

\bibitem[{\citenamefont{Mart and Benhold}(2000)}]{mart}
\bibinfo{author}{\bibfnamefont{T.}~\bibnamefont{Mart}} \bibnamefont{and}
  \bibinfo{author}{\bibfnamefont{C.}~\bibnamefont{Benhold}},
  \bibinfo{journal}{\prc} \textbf{\bibinfo{volume}{66}},
  \bibinfo{pages}{012201} (\bibinfo{year}{2000}).

\bibitem[{SAI()}]{SAID}
\emph{\bibinfo{title}{SAID partial wave analysis facility}},
  \urlprefix\url{http://gwdac.phys.gwu.edu/}.

\bibitem[{\citenamefont{Arndt et~al.}(2002)}]{arndt}
\bibinfo{author}{\bibfnamefont{R.~A.} \bibnamefont{Arndt}}
  \bibnamefont{et~al.}, \bibinfo{journal}{\prc} \textbf{\bibinfo{volume}{66}},
  \bibinfo{pages}{055213} (\bibinfo{year}{2002}).

\bibitem[{\citenamefont{Juli\'{a}-D\'{i}az et~al.}(2006)}]{bruno}
\bibinfo{author}{\bibfnamefont{B.}~\bibnamefont{Juli\'{a}-D\'{i}az}}
  \bibnamefont{et~al.}, \bibinfo{journal}{\prc} \textbf{\bibinfo{volume}{73}},
  \bibinfo{pages}{055204} (\bibinfo{year}{2006}).

\bibitem[{\citenamefont{Paterson}(2008)}]{paterson}
\bibinfo{author}{\bibfnamefont{C.~A.} \bibnamefont{Paterson}}, Ph.D. thesis,
  \bibinfo{school}{University of Glasgow} (\bibinfo{year}{2008}),
  \bibinfo{note}{{\sc CLAS} Collaboration, analysis of g8 data}.

\end{thebibliography}

\end{document}